\documentclass[11pt,a4paper]{article}

\usepackage{bbm,amsmath,amssymb,graphicx,array,amsfonts,cite,setspace,hyperref}

\usepackage{authblk,cite}
\usepackage[margin=1in]{geometry}

\title{Global and local epidemiology of Group A \textit{Streptococcus} indicates that naturally-acquired immunity is enduring and strain-specific} 
 \date{}

\author[a,b,1]{Rebecca H. Chisholm}
\author[c]{Jake A. Lacey} 
\author[d]{Jan Kokko}
\author[b,e]{Patricia T. Campbell}
\author[f]{Malcolm I. McDonald}
\author[g,h,i]{Jukka Corander}
\author[j]{Mark R. Davies}
\author[c,k,l]{Steven Y. C. Tong}
\author[b,e]{Jodie McVernon}
\author[e,m]{Nicholas Geard}

\affil[a]{School of Engineering and Mathematical Sciences, La Trobe University, Bundoora VIC, Australia}
\affil[b]{Centre for Epidemiology and Biostatistics, Melbourne School of Population and Global Health, The University of Melbourne, Melbourne VIC, Australia}
\affil[c]{Doherty Department, University of Melbourne at the Peter Doherty Institute for Infection and Immunity, Melbourne VIC, Australia}
\affil[d]{Department of Mathematics and Statistics, University of Helsinki}
\affil[e]{Victorian Infectious Diseases Reference Laboratory Epidemiology, University of Melbourne at the Peter Doherty Institute for Infection and Immunity, Melbourne VIC, Australia}
\affil[f]{Centre for Chronic Disease Prevention, Cairns Campus, JamesCook University, Cairns, Queensland}
\affil[g]{Helsinki Institute for Information Technology, Department of Mathematics and Statistics, University of Helsinki, Helsinki, Finland}
\affil[h]{Department of Biostatistics, University of Oslo, Oslo, Norway}
\affil[i]{Parasites and Microbes, Wellcome Sanger Institute, Wellcome Genome Campus, Hinxton, Cambridge, UK}
\affil[j]{Department of Microbiology and Immunology, University of Melbourne at the Peter Doherty Institute for Infection and Immunity, Melbourne VIC, Australia}
\affil[k]{Victorian Infectious Diseases Service, The Royal Melbourne Hospital}
\affil[l]{Global and Tropical Health Division, Menzies School of Health Research, Charles Darwin University, Darwin NT, Australia}
\affil[m]{School of Computing and Information Systems, Melbourne School of Engineering, The University of Melbourne, Melbourne VIC, Australia}

\begin{document}

\maketitle

\begin{abstract}
The bacterium Group A \emph{Streptococcus} (\textit{Streptococcus pyogenes},  GAS) is a human-specific pathogen and a major cause of global morbidity and mortality. Despite decades of research our knowledge of GAS infection and immunity is incomplete, hampering vaccine design and other efforts to reduce disease prevalence. Epidemiological studies indicate positive associations between the prevalence of GAS-related disease, the diversity of circulating strains and the degree of poverty in host populations. However, the infection and immune mechanisms underlying these associations are not clear.  
In this work, we use an agent-based model to demonstrate that observed diversity and prevalence are best accounted for by the hypothesis that GAS infection confers enduring strain-specific immunity, with reduced or absent cross-protection against infection by other strains. Our results suggest that the success of GAS vaccines will depend on their ability to elicit long-lasting cross-protective immunity over multiple strain types.
\end{abstract}

Group A \emph{Streptococcus} (\textit{Streptococcus pyogenes}, GAS) is a globally diverse pathogen that is responsible for a range of clinical presentations 
-- from superficial infections of the skin and throat, to cellulitis, life-threatening invasive infections, and post-infection sequelae including acute rheumatic fever (ARF), rheumatic heart disease (RHD), and acute post-streptococcal glomerulonephritis (APSGN)~\cite{cunningham2000pathogenesis}.  Superficial GAS infection accounts for the majority of GAS disease with an estimated 616 million incident cases of GAS pharyngitis occurring globally each year~\cite{carapetis2005global}.  
Severe GAS diseases are estimated to cause at least 500,000 deaths each year~\cite{carapetis2005global}.  
Of these, RHD is responsible for the greatest burden of disease, causing an estimated 319,400 deaths in 2015~\cite{watkins2017global}.
Safe and effective vaccines against GAS are urgently needed to reduce this disease burden.  However, a number of key scientific barriers to vaccine development remain, including our incomplete understanding of naturally acquired immune protection against re-infection and disease~\cite{Steer2016}.  Such understanding is crucial for determining the feasibility of a globally efficacious vaccine, and for the future evaluation of vaccine efficacy. 

Epidemiological studies present conflicting evidence for naturally acquired immunity. One recent analysis of data from a five-year longitudinal cohort study of GAS pharyngitis among school-age children in Pittsburgh, USA, found evidence for naturally acquired protection against the development of respiratory symptoms amongst children acquiring a previously detected strain, and partially heterologous strains~\cite{lewnard2020naturally}.  However, analysis of longitudinal data of GAS infection among Fijian school-age children over a 10-month period came to the opposite conclusion, finding no evidence that past exposure to particular strains or heterologous strains affected future GAS infections~\cite{campbell2020longitudinal}.  Such studies are limited by their inability to gauge prior GAS exposure and immunity in study cohorts~\cite{tsoi2015correlates}.  A murine challenge study showed evidence that sterilising strain-specific and enduring immunity requires two skin infections by the same GAS strain within a short-time frame~\cite{pandey2016streptococcal}.  Although this model of immune development has not been directly assessed in humans, we have previously shown that it can lead to patterns of transmission that are consistent with GAS epidemiology in at least one hyper-endemic population~\cite{chisholm2019epidemiological}.

\begin{figure}
\centering
\includegraphics[width=0.75\linewidth]{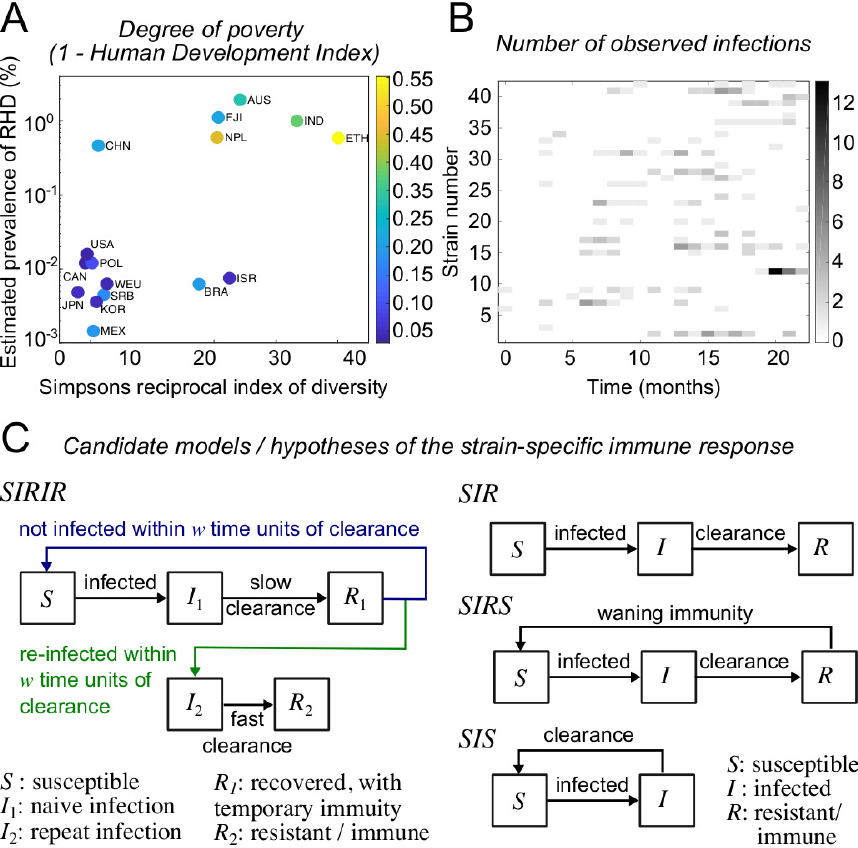}
\caption{Global and local GAS epidemiological data.  (A) Distribution of the (Simpson's reciprocal index of) diversity of GAS \emph{emm} types observed over, on average, a 10-year period~\cite{smeesters2009differences}, the estimated prevalence of RHD in all age groups~\cite{parnaby2010rheumatic,watkins2017global}, and (data point colour) the complement of the 2015 Human Development Index (HDI)~\cite{cooke2007indigenous,hdi} of populations in different countries/regions.  The HDI is a quality-of-life and standard-of-living index. AUS refers to Australian Aboriginal and Torres Strait Islander populations, and WEU to Western Europe.  All other labels correspond to ISO country codes. (B) Longitudinal data of GAS infection collected in one remote Australian Aboriginal community~\cite{mcdonald2008dynamic}. Here, the number of observed infections of each strain (\textit{emm} sequence type) is shown for each month of the study. (C) Candidate models of the immune response to GAS infection with respect to a single strain -- of $SIRIR$-type (enduring strain-specific immunity requires two infections by the same strain in quick succession~\cite{pandey2016streptococcal,chisholm2019epidemiological}), $SIR$-type (one infection leads to enduring immunity~\cite{gupta1998chaos-persisten}), $SIRS$-type (one infection leads to temporary immunity~\cite{buckee2004the-effects-of-}), and $SIS$-type (infection does not lead to immunity).}
\label{fig:fig1}
\end{figure}

A major challenge for understanding whether natural exposure to GAS confers protective immunity is the high diversity of circulating strains, and the complex epidemiology of GAS-related disease which varies markedly across different populations. Globally, over 240 different molecular sequence types (defined by the \textit{emm} sequence, which is a hypervariable region of the locus encoding the M-protein, a cell-surface protein common to all GAS isolates) of GAS have been identified to date, with an additional 1200 distinct allelic forms or sub-types~\cite{bessen2015molecular,bessen2019molecular}.
Meta-analyses of population-based studies~\cite{smeesters2009differences,parnaby2010rheumatic,watkins2017global} indicate that resource-poor populations tend to have both a higher number of GAS strains in circulation and a higher burden of RHD, compared to more affluent populations (Fig.~\ref{fig:fig1}A).  Epidemiological data on the prevalence of other GAS-related diseases, particularly superficial infections, is scarce~\cite{bowen2015global}.  However, given that RHD normally follows superficial GAS infection~\cite{cunningham2000pathogenesis}, the available data on RHD suggests a positive association between the prevalence of GAS-related disease more generally, strain diversity and the degree of poverty in a host population.

In Australia, the positive association between disease prevalence, strain diversity and poverty is evident when comparing GAS epidemiology between different sub-populations.  Remote Aboriginal and Torres Strait Islander communities account for 94\% of reported ARF cases~\cite{AIH2019}. Children in remote Aboriginal and Torres Strait Islander communities in northern Australia have some of the highest recorded prevalences of GAS-related skin infections, ARF and RHD globally~\cite{carapetis2005global,parnaby2010rheumatic,bowen2015global,watkins2017global}.  Studies conducted in these communities report dozens of co-circulating \textit{emm} types, which tend to move sequentially through communities and persist for a number of months~\cite{bessen2000contrasting,mcdonald2008dynamic}. In contrast, in urban settings such as the city of Melbourne, studies typically report only a few predominant \textit{emm} types, at much lower prevalence, causing mostly mild throat infections~\cite{rogers2007strain}.

Although there is currently no consensus on the immune response to GAS infection, it is likely contributing to the observed relationship between disease prevalence, strain diversity and population context.
Mathematical modelling studies of other multi-strain infectious disease systems have provided valuable insight into how the strain-specific immune response relates to epidemiology, particularly strain diversity~\cite{gupta1998chaos-persisten,dawes2002the-onset-of-os,gomes2002on-the-determin,chisholm2019epidemiological}.  However, existing models have not addressed the interaction between immunity, diversity, prevalence and population context.  Hence, the characteristics of the immune response that underlie the marked variation in GAS epidemiology across different population settings remain unknown.

In this study, we use an agent-based mathematical model of multi-strain GAS transmission together with likelihood-free inference methods, and local and global epidemiological data sets, to identify characteristics of GAS immunity. We use the model to compare four alternative hypotheses about the strain-specific immune response (shown schematically in Fig.~\ref{fig:fig1}C) and determine that global observations of GAS are best  accounted  for by a model  which assumes a single infection confers enduring strain-specific immunity,  with  reduced  or  absent  cross-protection  against  infection  by other strains.  We validate this finding by calibrating the model to longitudinal population data ~\cite{mcdonald2008dynamic} (Fig.~\ref{fig:fig1}B) of a specific host setting -- a remote Aboriginal community in northern Australia -- which, in addition, enables us to obtain, for the first time, an estimate of a basic reproduction number ($\mathcal{R}_0$) for GAS. 

\section*{Results}

Our agent-based model allowed us to compare three alternate hypotheses for the immune response to GAS infection: an $SIR$-type response~\cite{gupta1998chaos-persisten} (where one infection leads to permanent immunity), an $SIRS$-type response~\cite{buckee2004the-effects-of-} (where one infection leads to temporary immunity), and an $SIS$-type response (where infection does not lead to protective immunity).  We used a previously proposed model to explore a fourth hypothesis for the immune response: an $SIRIR$-type response~\cite{chisholm2019epidemiological,pandey2016streptococcal} (where enduring immunity requires two skin infections by the same GAS strain within a short-time frame).
We conducted \emph{in silico} experiments to 
identify infection- and immune-parameter regions in both models where changes to population-specific parameters, specifically, the basic reproduction number $\mathcal{R}_0$ and population size $N$, were sufficient to explain global epidemiological trends of GAS.  The alternative model formulations explored are set out in SI Appendix, Table~S1. 
\subsection*{Endemic diversity and prevalence are greatest when there is low cross-protective immunity} 
The immune response to GAS must be able to support populations with a medium--high prevalence of infection, $P(t)$ -- calculated as the percentage of hosts in the population that are infected by at least one strain at a particular point in time $t$ -- with high 10-year period strain diversity, $D(t)$ (which is a measure of the total number of strains as well as how evenly strains are distributed across all infections in the population over a ten year period), henceforth referred to as `diversity' (Fig.~\ref{fig:fig1}A).  
We first searched for infection- and immune-parameter regions in our model 
where it was possible for populations to experience endemic transmission with a medium--high $P(t)$ and high diversity $D(t)$. 

We considered four sets of infection and immunity scenarios, with each set defined by the assumed mean duration of immunity $1/\theta$ (6 months or 5 years), and level of resistance of infected hosts to acquiring additional infections (the level of resistance of hosts to co-infection) $x$ (low or high, corresponding to ${x=10}$ or ${x=100}$). For each of these scenario sets, we varied the strength of strain-specific immunity $\sigma$ (where $0\le\sigma\le1$) and cross-protective (heterologous, or cross-strain) immunity $\omega$ (where $0\le\omega\le\sigma$) over their full ranges, and kept all other model parameters fixed at their base values (as shown in SI Appendix, Table~S1).  Scenarios where ${\sigma=\omega=0}$ corresponded to an $SIS$-type immune response; all other scenarios corresponded to an $SIRS$-type immune response.  We first assumed that strain-specific and cross-strain immunity acted by reducing the susceptibility of hosts to subsequent infection by the same or different strains (example model outputs from single simulations of this model are shown in SI Appendix, Fig.~S1.)  We then re-ran all simulations using an alternative model formulation where we assumed an additional effect of cross-strain immunity -- that it also reduced the duration of infections by strains that were present in a host when the host cleared another strain.   

\begin{figure}
\centering
\includegraphics[width=0.7\linewidth]{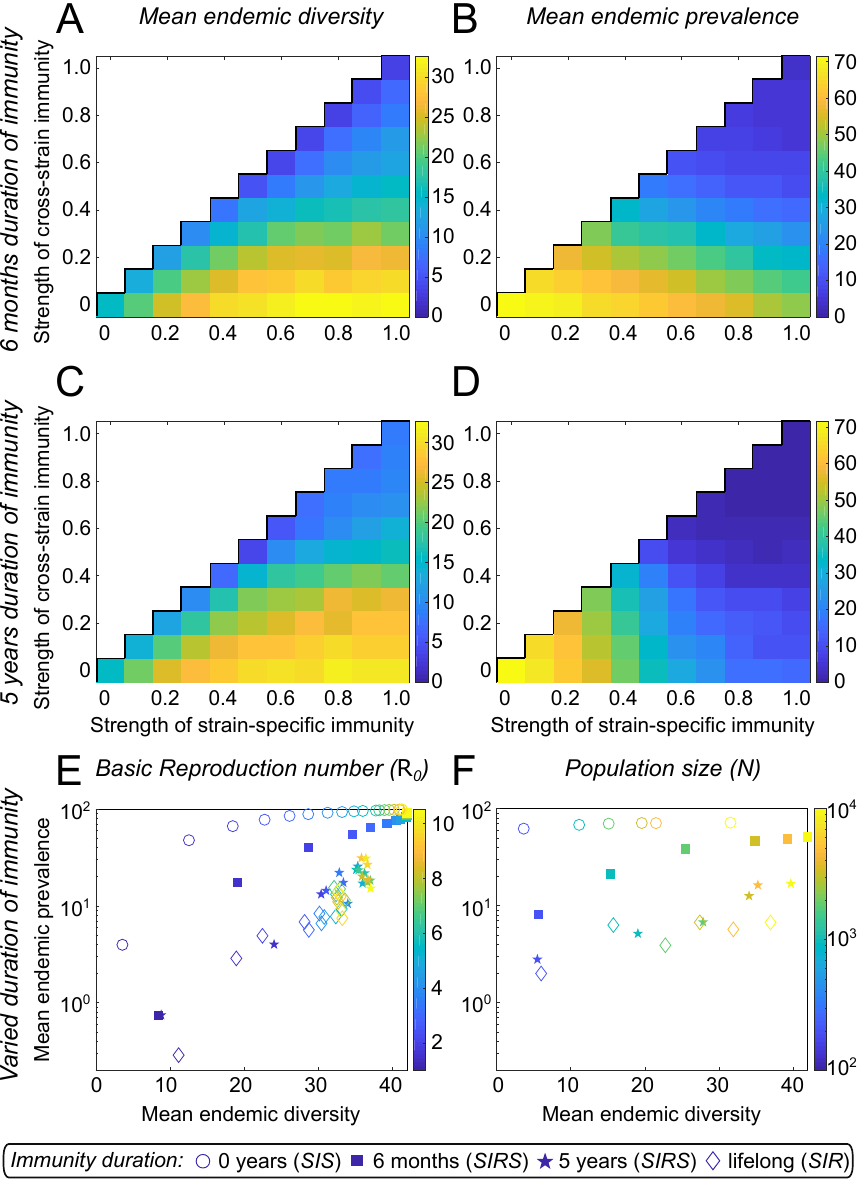}
\caption{The mean endemic (A,C) diversity of strains  $D(t)$, and (B,D) prevalence of infections $P(t)$ for different strengths of strain-specific immunity $\sigma$ (horizontal axes) and cross-strain immunity $\omega\le\sigma$ (vertical axes), when there is low resistance to co-infection, and the mean duration of immunity $1/\theta$ is (A,B) 6 months; and (C,D) 5 years.  Comparing epidemiological dynamics across populations characterised by different values (as indicated by the colour of points) of (E) the basic reproduction number $\mathcal{R}_0$; and (F) the population size $N$, and under different assumptions about $1/\theta$ (indicated by the shape of points). The position of points indicates a population's mean endemic $P(t)$ (vertical axes) and associated mean endemic $D(t)$ (horizontal axes).  Here,  (A--D) ${\mathcal{R}_0=2.07}$,  ${N=2500}$, ${\sigma=0,0.1,\hdots,1}$, ${\omega=0,0.1,\hdots,1}$; (E--F)  ${\omega=0.1}$, ${\sigma=1}$, $1/\theta=0,0.5,5,71$ (years); (E) ${N=2500}$, $\mathcal{R}_0=1,1.5, \hdots, 10.5$;  (F) ${\mathcal{R}_0=2.07}$, $N=0.2,1,2,3.5,5,10$ ($10^3$ hosts). Other parameter values are given in the SI Appendix.}
\label{fig:fig2}
\end{figure}

In both formulations of our model, we found that the set of within-host parameters which led to the highest mean endemic strain diversity $D(t)$ did not correspond to those which maximised the mean endemic prevalence $P(t)$.  However, an intermediate mean endemic $P(t)$ and high mean endemic $D(t)$ could be achieved when there was weak cross-strain immunity and an intermediate--high strength of strain-specific immunity. 
Specifically, when comparing the mean endemic $P(t)$ across transmission scenarios, it was always lower in scenarios defined by a higher value of any of the parameters controlling immunity or the level of resistance to co-infection ($\sigma$, $\omega$, $1/\theta$, and $x$) (Fig.~\ref{fig:fig2}B,D and SI Appendix, Fig.~S2B,D and Fig.~\ref{fig:figs6}C,D).    The corresponding differences in the mean endemic $D(t)$ were more varied (Fig.~\ref{fig:fig2}A,C and SI Appendix, Fig.~S2A,C, Fig.~S3, Fig.~\ref{fig:figs6}A,B  and Fig.~\ref{fig:figs7}A,B).  Across scenarios with a higher level of resistance to co-infection, $x$, or a higher strength of cross-strain immunity, $\omega$, the mean endemic $D(t)$ was always comparatively lower.  However, the difference in the endemic $D(t)$ across scenarios defined by either a longer mean duration of immunity $1/\theta$ or higher strength of strain-specific immunity $\sigma$, was either positive or negative. 

 \subsection*{Long-lived strain-specific immunity conferred by a single infection best reproduces global epidemiological trends}
Within the narrower parameter space in our model where it was possible to achieve an intermediate mean endemic $P(t)$ with a high mean endemic $D(t)$, we then searched for parameter regions where a positive association between prevalence and diversity existed (among populations characterised by either a different $\mathcal{R}_0$ or population size $N$) that resembled global GAS epidemiology. 
We considered eight sets of scenarios, each defined by a different mean duration of immunity (${1/\theta\in\{0, 0.5, 5,71\}}$ years) and level of resistance to co-infection (${x\in\{10,100\}}$).  Scenarios with $1/\theta=0$ and 71 years corresponded, respectively, to an $SIS$ and $SIR$-type model of the immune response, while $1/\theta=0.5$ and 5 years corresponded to an $SIRS$-type model of the immune response.  In all scenarios, we assumed immunity to be highly strain specific ($\sigma=1$, $\omega=0.1$). For each scenario, transmission was first simulated in populations of size $N=2500$ with $\mathcal{R}_0$ ranging from ${1\le\mathcal{R}_0\le10.5}$ ($\mathcal{R}_0$ was varied between populations by varying the expected number of host contacts $c$), keeping all other parameters fixed.  We then simulated transmission in populations with $N$ ranging from ${200\le N\le10,000}$ and with $\mathcal{R}_0=2.07$. 

In scenarios where the mean duration of immunity $1/\theta$ was short (zero or six months), populations with a higher value of $\mathcal{R}_0$ always had a higher mean endemic prevalence $P(t)$ and strain diversity $D(t)$ compared to populations with lower $\mathcal{R}_0$ (Fig.~\ref{fig:fig2}E and SI Appendix, Fig.~S4A and Fig.~S6E). This trend was independent of the level of resistance to co-infection, and whether or not our model accounted for the additional effect of cross-strain immunity on co-infected hosts.  Population size $N$ had a similar effect on mean endemic $D(t)$; mean endemic $P(t)$, however, only increased for larger populations when $1/\theta>0$ (Fig.~\ref{fig:fig2}F and SI Appendix, Fig.~\ref{fig:figs4}B and Fig.~\ref{fig:figs6}F). In scenarios with a longer mean duration of immunity (5 years and above), high levels of mean endemic $D(t)$ could be achieved with lower, and arguably more realistic, levels of mean endemic $P(t)$~\cite{mcdonald2008dynamic}.  In these enduring-immunity scenarios,  the mean endemic $D(t)$ and $P(t)$ were highly sensitive to a population's basic reproduction number $\mathcal{R}_0$ and size $N$ when they were sufficiently small ($\mathcal{R}_0\lesssim2.5$ and $N\lesssim3000$).  Above these thresholds, populations with a higher value of $\mathcal{R}_0$ or $N$ resulted in similar mean endemic $D(t)$ and $P(t)$. 

To understand why $D(t)$ and $P(t)$ were insensitive to $\mathcal{R}_0$ and $N$ when $1/\theta$ was sufficiently large, we ran additional simulations with $1/\theta$ set at values ranging from 5 to 11 years and investigated the relationship between $1/\theta$ and the time until re-infection by the same strain for individual hosts. There was substantial overlap between distributions of the time until re-infection by the same strain for all values of $1/\theta$ considered, suggesting there is little difference in the overall transmission dynamics in our model once $1/\theta$ is sufficiently large (SI Appendix, Fig.~\ref{fig:figs4}C and Fig.~\ref{fig:figs7}C). 

The analogous parameter exploration was then repeated using the $SIRIR$-type model described in~\cite{chisholm2019epidemiological}, which assumes that lasting strain-specific immunity to GAS requires two infections by the same strain within a specific time interval, the so-called \textit{inter-infection interval}, $w$. In this analysis, scenario sets (within which $\mathcal{R}_0$ and $N$ were varied, as above) were characterised by a specific value of the inter-infection interval (3 or 19 weeks, which were determined to be possible values in~\cite{chisholm2019epidemiological}) and level of resistance to co-infection (${x\in\{10,100\}}$).  Both the mean endemic $D(t)$ and $P(t)$ were insensitive to changes in $\mathcal{R}_0$ (as well as $x$ and $w$), but both were found to be increasing functions of $N$ (SI Appendix, Fig.~\ref{fig:figs5}).


\subsection*{Model calibration to a GAS endemic community indicates that infection confers long-lived strain-specific immunity}

Last, we calibrated our model (which best explained global epidemiological trends) to a particular host setting using longitudinal data (Fig.~\ref{fig:fig1}B) collected in a remote Aboriginal community in northern Australia where GAS disease is hyper-endemic~\cite{mcdonald2008dynamic}. In this previous study, data was collected via prospective surveillance of a population of approximately 2500 people, monthly over a 23-month period. Swabs were taken from the throats of all participants and any skin sores of participants and GAS isolates were \textit{emm} sequence and pattern-typed.  Between 1--11 \emph{emm} types were found to be circulating in the community at any one time-point for variable durations (see \cite{mcdonald2008dynamic} for full details and analysis). 

To fit our model to this data, we used the Python implementation of the Likelihood-Free Inference by Ratio Estimation approach, PYLFIRE, which utilises logistic regression based classification to approximate a posterior distribution~\cite{kokko2019pylfire}. We ran this inference method with three unknown parameters: $\mathcal{R}_0$ (which is likely to be specific to particular host settings), as well as the strength of strain-specific immunity $\sigma$ and mean duration of immunity $1/\theta$ (which we assumed to be common among settings).  Although, we did not expect that $1/\theta$ would be identifiable, given the insensitivity of the epidemiological dynamics to large values of $1/\theta$ (Fig.~\ref{fig:fig2}E,F). All other immunity parameters were fixed at values consistent with the results of the model exploration (\emph{i.e.}, ${\omega=0.1}$), and the level of resistance to co-infection was assumed to be low (${x=10}$) since  GAS was isolated from multiple sites in participants (from skin sores and throat). Population parameters remained unchanged from those in our base model, which were originally chosen to be consistent with this population setting.   Further details of the fitting approach, including our two choices of prior distributions for the free parameters, are provided in SI Appendix.

\begin{figure}
\centering
\includegraphics[width=0.75\linewidth]{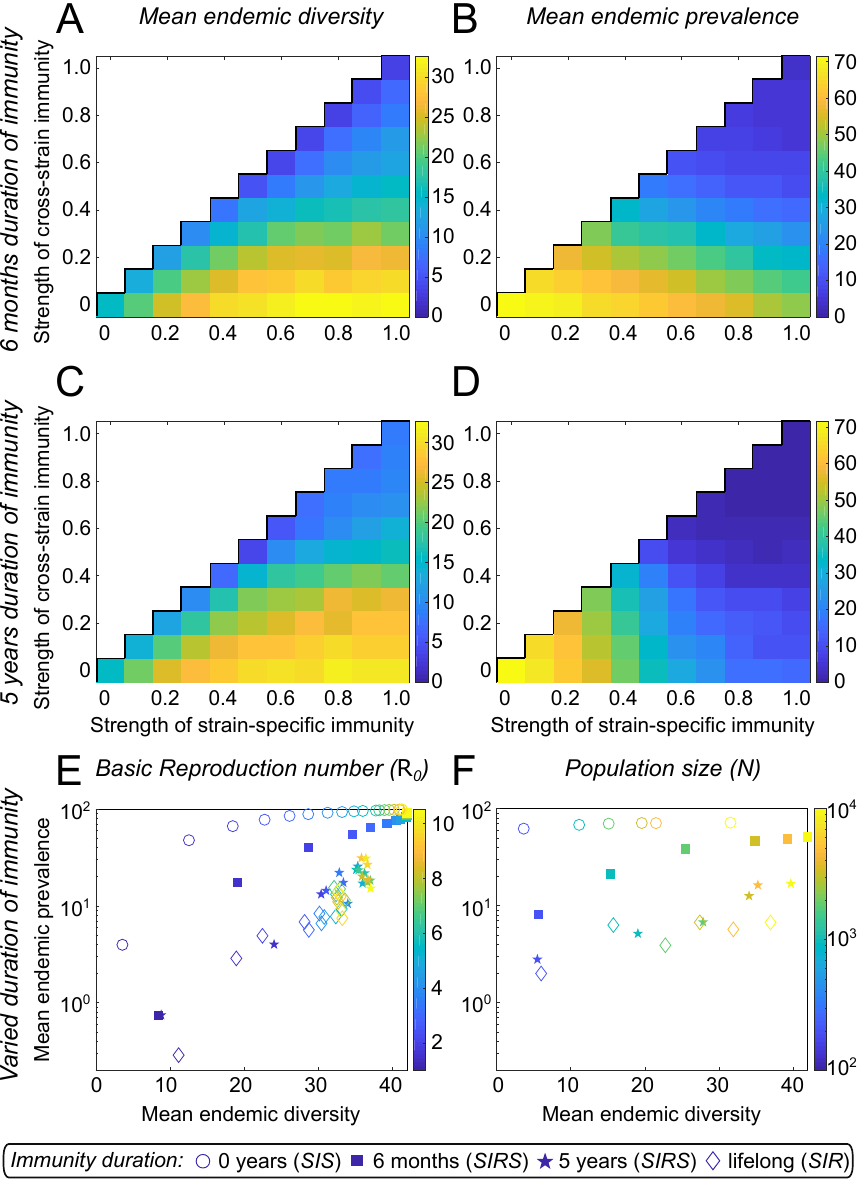}
\caption{Model calibration to longitudinal data from a single population assuming mean duration of immunity is between 1--15 years. Estimated (A) marginal distribution for $\mathcal{R}_0$; (B) marginal distribution for $\sigma$; and (C) joint-marginal distribution for $\mathcal{R}_0$ and $\sigma$. Fixed parameter values are given in the SI Appendix. }
\label{fig:fig3}
\end{figure}

Both choices of priors led to similar posterior distributions for $\mathcal{R}_0$ and $\sigma$ (compare Fig.~\ref{fig:fig3} and SI Appendix, Fig.~\ref{fig:figs8}).
For the first case, the estimated mean value of $\mathcal{R}_0$ was 2.27 (50\% CI: 1.27, 3.22) and for $\sigma$ was 0.78 (50\% CI: 0.66,0.90). For the second case, the estimated mean value of $\mathcal{R}_0$ was 2.19 (50\% CI: 1.15, 3.15) and for $\sigma$ was 0.78 (50\% CI: 0.67,0.91).  The median, mode and 95\% credible intervals for $\mathcal{R}_0$ and $\sigma$ for both cases are provided in SI Appendix, Table~S2.  Our results suggest that the $\mathcal{R}_0$ of GAS in this population setting is comparable to previous estimates of $\mathcal{R}_0$ for \emph{Streptococcus pneumoniae}~\cite{hoti2009outbreaks,gjini2017geographic} and \emph{Staphylococcus aureus}~\cite{hogea2014basic} (which occupy similar ecological niches to GAS) and are consistent with our findings from the parameter exploration described above that the immune response to GAS infection is enduring and highly strain-specific

\section*{Discussion}

Our knowledge of immune response to GAS infection is incomplete. In this study we used an agent-based model of GAS transmission to gain mechanistic insight into the immune response to GAS infection using population-level observations of strain diversity and infection prevalence from different host settings. 


\subsection*{A hypothesis for the immune response to GAS infection}

Our analysis supports the hypothesis that GAS is characterised  by infection which confers an intermediate to high level of enduring strain-specific immunity to re-infection, with reduced or absent cross protection against infection by other strains (i.e., an $SIRS$- or $SIR$-type model of the strain-specific immune response). 
Under these assumptions, we demonstrated how varying the population-specific parameters, $\mathcal{R}_0$ and population size $N$, can shift GAS epidemiology from a low prevalence, low diversity regime to a high prevalence, high diversity regime, consistent with observed global patterns~\cite{carapetis2005global,smeesters2009differences,bowen2015global}. 
The sensitivity of strain diversity to variation in $\mathcal{R}_0\lesssim2.5$, 
along with known limitations in data availability~\cite{watkins2017global}, might also help to understand some of the outlier population groups, such as Brazil and Israel, that have high recorded strain diversity with relatively low reported prevalence of GAS disease (Fig.~\ref{fig:fig1}A).  
This model also captures patterns of transmission at the scale of a single community.  

This hypothesis is consistent with other epidemiological data  which clearly show 
that infection is much less frequent in adults than in  children~\cite{mcdonald2008dynamic,bowen2015global}, (indicating enduring strain-specific immunity can be acquired), as well as
individual hosts being infected by the same strain at multiple times~\cite{mcdonald2008dynamic,bessen2000contrasting} (indicating imperfect strain-specific immunity), by 
different strains at multiple times~\cite{bassett1972streptococcal} (indicating imperfect cross-strain immunity), and by multiple strains at the one time (indicating the possibility of mixed-strain co-infections)~\cite{Carapetis1995,bessen2000contrasting}.  Our analysis enables these disparate empirical results to be synthesised into a coherent dynamic framework for the first time, providing a unified account of the relationship between the immune response in individuals and observed population-level dynamics of GAS. 

\subsection*{The interaction between the immune response to infection, prevalence and strain diversity}

Our results provide insight relevant for multi-strain pathogen transmission in general, particularly in relation to the effects of co-infection and duration of immunity (discussed below).  Insights gained into the effects of strain-specific and cross-strain immunity on period strain diversity are consistent with findings on the relationship between these parameters and instantaneous strain diversity (or strain coexistence) from other multi-strain pathogen modelling studies~\cite{gupta1998chaos-persisten,dawes2002the-onset-of-os,gomes2002on-the-determin,buckee2004the-effects-of-}. 

We found that the tendency for a pathogen to cause co-infection can counteract the loss of strain diversity that arises when clearance of infection does not confer strain-specific immunity.  When clearance of infection does confer strain-specific immunity, high levels of strain diversity (both instantaneous and over time) are possible, even though the overall prevalence of infection is reduced by this immunity.  
We found that this effect of strain-specific immunity on strain diversity over time can be magnified for pathogens with a longer duration of immunity (if the the strength of cross-strain immunity is sufficiently high), even though the prevalence of infection is further reduced.

\subsection*{Implications for control of GAS}

Our findings suggest that the high strain diversity that characterises hyper-endemic settings is very stable, particularly in disadvantaged populations where systemic social issues likely contribute to high rates of transmission~\cite{mcdonald2008dynamic,bessen2000contrasting}.   
Nevertheless, interventions that are capable of reducing strain diversity could have a significant impact on the prevalence of GAS infections.  For example, control measures capable of increasing the level of cross-protection conferred by infection, such as a vaccine that broadens the immune response following natural infection~\cite{pandey2018skin}, may lead to sustained reductions in the diversity of GAS strains circulating in host populations, and thus also to prevalence. 

Reducing the frequency of co-infection could also substantially reduce diversity and prevalence.  Shorter infections have a greater chance of resolving prior to hosts acquiring additional infections compared to longer infections.  Therefore, interventions that lead to a shorter mean duration of infection should, in turn, reduce the frequency of co-infection.  In the context of disadvantaged populations, empowering people with education about the importance of rapid treatment, removing barriers to accessing healthcare, and increasing surveillance would be expected  to reduce the frequency of co-infection as well as prevalence and diversity.

\subsection*{Model limitations}
To enable an extensive exploration of the complex dynamics of infection and immunity in a multi-strain system, our model makes several simplifying assumptions. 
Our model does not include strain evolution which is likely, but not yet quantified, in hyper-endemic settings
where co-infection provides opportunities for recombination~\cite{mcgregor2004multilocus}.  Despite lacking explicit strain evolution, our model captures the consequences of such a process by modelling the importation of novel strains into a population via migration.  
Our model does not include differences in infection type in relation to the presence or absence of symptoms (carriage versus symptomatic infection) or in the location of infection (skin versus throat).  It is unknown whether the immune response following throat infection, throat carriage, and skin infection are comparable, nor how the duration and infectiousness of these infection types compare.  We have made the most-parsimonious assumption by treating all infection types as the same. 
We assume that strains have equal fitness. With this assumption, it is the possibility of co-infection and/or the conferral of enduring strain-specific immunity following infection clearance that enables the transmission of many different strains over time.  However, a recent study of \emph{S. pneumoniae} has shown that selection acting on accessory genes that are more advantageous to the bacteria when they are rarer makes it possible for strains with substantial fitness differences to coexist~\cite{corander2017frequency}. 
We also assume that an infection by any strain is equally likely to lead to a more serious episode of GAS disease.  However, we know that so-called rheumatogenic and nephritogenic strains have been associated, respectively, with ARF/RHD and APSGN in some populations, although these aren't always consistent across populations~\cite{metzgar2011m}.  
Finally, we have not included heterogeneity in host population structure.  However, households in particular are thought to be important sites of transmission for GAS skin infections in hyper-endemic settings~\cite{bowen2016whole}, and household structure can differ significantly between settings of poverty and more-affluent settings~\cite{vino2017indigenous}.  
Future studies could investigate whether 
the addition of these types of extra pathogen or host-population complexities into our model would alter our conclusions regarding  the immune response to GAS infection.

\section*{Methods}
\subsection*{Model of GAS transmission}
Our agent-based model is based on the $SIRIR$ GAS multi-strain model proposed in \cite{chisholm2019epidemiological} (where it is fully described) with a modified model of the immune response to infection. The model is implemented in \textsc{Matlab}. %
Briefly, the  model simulates the transmission of $n(t)$ strains of GAS in a well-mixed host population (where agents correspond to hosts) of constant size $N$ by explicitly modelling the outcome of contact events between hosts (which occur at mean rate of $c$ contacts per day). The model incorporates the birth and death of hosts (at the per capita rate $d$), and host migration into and out of the population (at the per capita rate $\alpha$).   Hosts can be infected with up to $\kappa\in\mathbb{Z}^+$ infections at any time, which can include multiple infections of the same strain and infections by multiple strains.   
The susceptibility of hosts to infection decreases by a factor $(1-\sum_jG_{i,j}(t)/\kappa)^x$ (where $G_{i,j}(t)$ is the number of infections of strain $j$ in host $i$ at time $t$, and $x$ is the a number scaling the level of resistance to acquisition of new infections) as the total number of infections in each host increases. This assumption incorporates the effect of pathogen populations directly competing for space and resources within the host, or indirectly interacting via the host immune response.  
Strains prompt a distinct immune response in hosts, cause infections with the same mean duration of infection $1/\gamma$, and have identical transmissibility $\beta$ (the base probability of an effective contact), which leads to a strain-specific \emph{basic reproduction number}, $\mathcal{R}_0$ (the expected number of secondary infections caused by a single infected host introduced into a completely susceptible host population), which is the same as that of the overall pathogen.  In our model, ${\mathcal{R}_0=c\beta/(\gamma+d+\alpha)}$.

\subsubsection*{Model of the immune response to infection}
We assume that clearance of a strain confers strain-specific immunity of strength ${0\le\sigma\le1}$ and immunity to all other strains (cross-strain immunity) of strength ${0\le\omega\le\sigma}$.  Immunity further reduces the susceptibility of hosts to subsequent infection by the same and different strains by a factor $(1-\sigma)$ and ($1-\omega$) respectively, and lasts for a mean duration of $1/\theta$.    In an alternative model, described in the SI Appendix, we also allow cross-strain immunity to reduce the duration of infections by strains that are present in a host when the host clears another strain (so that the new mean duration of these infections is $(1-\omega)/\gamma)$.
With this set-up, our model can explore (on average) an $SIRS$-type model of the immune response (when the mean duration of immunity $1/\theta$ is less than the lifetime of hosts and ${\sigma>0}$), an $SIS$-type model (if either ${\sigma=0}$ and/or ${1/\theta=0}$),  or an $SIR$-type model (when ${\sigma=1}$ and ${1/\theta}$ is greater than or equal to the lifetime of hosts).

\subsubsection*{Summary statistics} Previous GAS  epidemiological studies report the \emph{point prevalence} (\emph{i.e.}, instantaneous prevalence) or incidence rate of particular GAS diseases, such as pharyngitis~\cite{carapetis2005global}, impetigo~\cite{bowen2015global}, or RHD~\cite{watkins2017global} to describe the burden of disease in different geographic regions.   Data are also available on the frequency of circulating GAS strains over a period of time, and from which the \emph{period diversity} of GAS strains has been calculated~\cite{smeesters2009differences}.  To be consistent with these measures of disease burden, we use the following two metrics to describe GAS epidemiology simulated from our model: the population point prevalence of infection, $P(t)$, and the period diversity of strains, $D(t)$.  $P(t)$ is calculated as the percentage of hosts that are infected by at least one strain at a particular point in time, $t$, while
$D(t)$ is calculated as:
\begin{equation}\label{eq:diversity}
D(t)=\frac{M(t;u)(M(t;u)-1)}{\sum_{j}{m}_j(t;u)({m}_j(t;u)-1)},
\end{equation}
where ${{m}_j(t;u)=\sum_{v=t-u}^{t}\sum_i{G}_{i,j}(v)}$ is the number of infections of strain $j$ in the  population between times ${t-u}$ and $t$, and ${M(t;u)=\sum_j{m}_{j}(t;u)}$ is the total number of infections in the population between times ${t-u}$ and $t$. If $t<u$, then ${{m}_j(t;u)}$ is calculated as ${{m}_j(t;u)=\sum_{v=0}^{t}\sum_i{G}_{i,j}(v)}$. 
Period diversity is distinct from instantaneous diversity which is a measure of the diversity of strains in circulation at a given point in time. Instantaneous diversity is typically used to describe transmission in mathematical modelling studies of multi-strain pathogens.  A number of previous studies ~\cite{gupta1998chaos-persisten,dawes2002the-onset-of-os,gomes2002on-the-determin} have investigated the relationship between instantaneous strain diversity and cross-strain immunity, finding that a high level of this type of strain diversity can be maintained at endemicity in host populations when there is a low level of cross-strain immunity conferred by the clearance of individual strains, and also that high cross-strain immunity leads to low instantaneous diversity. 
Instantaneous diversity can be calculated using Eq.~\ref{eq:diversity} with ${{m}_j(t;u)=\sum_i{G}_{i,j}(t)}$.

\subsection*{Details of \emph{in silico} experiments}

Justification for the selection of base model parameters is provided in the SI Appendix.  \emph{In silico} experiments are run for 30 years to allow the epidemiological dynamics to reach equilibrium.  Unless otherwise stated, every model scenario defined by a unique combination of model parameters is simulated ten times to determine mean behaviour.  Further details of the model implementation are detailed in the SI Appendix.

\paragraph{Acknowledgments} We thank Jonathan Carapetis and Ross Andrews for contributions to data collection, and David Price and Brittany Rose for statistical advice. This work was supported in part by NHMRC project grants (GNT1098319 and GNT1130455) and NHMRC Centre of Research Excellence (GNT1058804). JK was supported by an Academy of Finland grant (316602), JC is supported by an ERC grant (742158), SYCT is supported by an NHMRC Career Development Fellowship (CDF1145033), JM is supported by an NHMRC Principal Research Fellowship (PRF1117140) and MRD is supported by a University of Melbourne C.R. Roper Fellowship


\renewcommand{\thefigure}{S\arabic{figure}}
\renewcommand{\theequation}{S\arabic{equation}}
\renewcommand{\thetable}{S\arabic{table}}

\setcounter{figure}{0}   
\setcounter{equation}{0}  
\setcounter{table}{0}

\newpage
\section*{SI Appendix}
\bigskip

\subsection*{Details of model implementation}
\subsubsection*{Time steps}
Model time steps have a length $\tau$ such that $t=k\tau$ at time-step ${k\in\mathbb{Z}}$.  In all results presented we assume a time-step length $\tau$ of one day.    

\subsubsection*{Infection}
We set $\kappa$ to be $n_{\rm max}$ so that, potentially (but very unlikely), a host may be co-infected with every strain in the population.  If the the level of resistance to co-infection $x\gg 10$, then co-infection is unlikely.  If $x$ is of the order of 10 or below, then co-infection is likely, but the chances of acquiring additional infections decreases as the number of co-infections approaches $\kappa$. 
We also assume that superinfection (where a new infection supersedes a current infection) is not possible in the model. Thus, a host at infection capacity must clear an infection before a new one can establish. 

At each time step, infections by each strain in a host clear with probability ${\Gamma = 1 - \exp(-\tau \gamma)}$, where $1/\gamma$ is the mean duration of a single infection.  A host with multiple infections of the same strain will clear all of these infections simultaneously.   Clearance of an infection confers strain-specific immunity of strength ${0\le\sigma\le1}$ and immunity to all other strains (cross-strain immunity) of strength ${0\le\omega\le\sigma}$.  At each time step, hosts may also lose their immunity to each strain independently with probability ${\Theta = 1 - \exp(-\tau\theta)}$, where $1/\theta$ is the mean duration of immunity.  

\subsubsection*{Transmission}
Each host contacts $C$ other hosts per day, where $C$ is a Poisson-distributed random variable with mean $c$ (the mean number of daily contacts).  
A host's susceptibility determines whether a contact event with an infectious host results in transmission.  
The susceptibility $Q_{i,j}(t)$ of host $i$ to infection by strain $j$ at time $t$ is calculated as ${Q}_{i,j}(t)=\beta r s$, where $\beta>0$ is the probability a completely susceptible host is infected during a contact event with an infected host, $r$ is the susceptibility of the host relative to an uninfected host, and $s$  is the susceptibility of the host relative to a host without immunity.

The scalar quantity $r$ incorporates the effects of strains directly competing for space and resources within the host, or indirectly interacting via the host immune response. It is calculated as
 \begin{equation}
 r=\Big(1-\sum_j{G}_{i,j}(t)/\kappa\Big)^x,
 \end{equation}
 where ${G}_{i,j}(t)\in\{0,1,2,\hdots,\kappa\}$, is the number of infections of strain $j$ carried by host $i$ at discrete time $t$, and $x>0$ is a number scaling the level of resistance of acquisition of new infections due to the competitive advantage of already established infections. 

The scalar quantity $s$, the susceptibility of the host relative to a host without immunity,  is calculated as
\begin{align}
s &=
\begin{cases}
1-\sigma,&\text{if }{H}_{i,j}(t)=1,\\
1-\omega,&\text{if } {H}_{i,j}(t)=0\text{ and } {\sum_k{H}_{i,k}(t)>0},\\
1,&\text{if } \sum_k{H}_{i,k}(t)=0,
\end{cases}
\end{align}
where ${H}_{i,j}(t)\in\{0,1\}$, takes a value of one if host $i$ has protective immunity to strain $j$ at discrete time $t$, and zero otherwise.  

\subsubsection*{Host contacts}
The $C$ contacts of each infectious host are chosen uniformly at random from the population, and the outcomes of these contact events are then determined (\textit{i.e.}, whether or not the infectious host infects a contact). We specify that transmission may only occur one-way from the infectious host to their contacts.   If the infectious host has multiple infections, one infection is chosen uniformly at random to attempt transmission.   If this attempt fails, then the contact event does not result in transmission. 

\subsubsection*{Migration}
To model migration, each day we select $A$ hosts (where $A$ is a Poisson distributed random variable with mean $\alpha N$) uniformly at random to be replaced by immigrants.  Immigrants are assumed to have no immunity to current strains
but may be infected with up to one copy of infection of any of the $n_{\rm max}$ strains circulating in the region.

\subsubsection*{Base model parameter selection}
The base values of the population size $N$, the maximum number of strains $n_{\rm max}$, and the death rate $d$ are set at 2500, 42 and 1/71 per year respectively to be consistent with community sizes~\cite{mcdonald2008dynamic}, the number of strains in circulation~\cite{smeesters2009differences}, and life expectancy~\cite{abs2018} in one particular hyper-endemic setting: Indigenous communities in northern Australia.  
Migration patterns are not described in these settings. Anecdotally, Indigenous Australians are described as having a higher than average rate of mobility compared to non-Indigenous Australians~\cite{morphy2007uncontained}.  
We set the per capita expected migration rate $\alpha$ to $1.71\times10^{-4}$ per day. With the prevalence of infection in migrants set to 10\%, on average, an infected migrant enters the population approximately once every three weeks, which is consistent with genomic analysis of GAS isolates collected across two Indigenous communities in Northern Australia~\cite{Marcato2019}. 

The basic reproduction number $\mathcal{R}_0$ for GAS is unknown.  We specify a base value of ${\mathcal{R}_0\approx2}$ in the model, consistent with estimates of $\mathcal{R}_0$ for other pathogenic bacteria that occupy similar niches to GAS: \emph{Streptococcus pneumoniae}~\cite{hoti2009outbreaks,gjini2017geographic} and \emph{Staphylococcus aureus}~\cite{hogea2014basic}.  
The mean duration of infection $1/\gamma$ is set at two weeks to be consistent with clinic data collected in hyper-endemic settings~\cite{mcdonald2008dynamic,bessen2000contrasting,mcdonald2007molecular}, while the base number of daily contacts $c$ is set at $c\approx5$ to be consistent with estimates of the mean number of daily physical contacts between individuals reported in European countries~\cite{mossong2008social}.  
For each combination of the epidemiological parameters $\{\mathcal{R}_0,1/\gamma,c,\alpha\}$ considered, the baseline transmission probability $\beta$ is calculated using Equation~(3). 

Since GAS can infect multiple sites in a host (including pharyngeal and skin infections)~\cite{Carapetis1995,bessen2000contrasting}, we expect there to be low resistance to GAS co-infection for at least two co-infections, and that resistance will increase quickly as the number of co-infections exceeds two.  We therefore set the baseline value of the level of resistance to co-infection to be ${x=10}$ to be consistent with this description of co-infection.  However, in order to explore the role of co-infection in shaping observed trends in GAS epidemiology,  we also repeat simulations for every model scenario considered with ${x=100}$, which corresponds to the assumption that co-infection is very unlikely.  

The duration of immunity to GAS in humans is unknown.  Large outbreaks of acute post-streptococcal glomerulonephritis linked to particular ``nephritogenic" strains in a hyper-endemic region were shown to occur approximately every 5 years~\cite{marshall2011acute}, suggesting immunity to these strains could last at least 5 years.   Therefore, we consider model scenarios where the mean duration of immunity $1/\theta$ is 5 years.  However, given the uncertainty in this parameter estimate, we also consider scenarios where immunity has a shorter mean duration of 6 months, when it is lifelong (mean duration of 71 years), as well as when immunity does not develop at all post infection (${1/\theta=0}$).  The strengths of strain-specific immunity $\sigma$ and cross-strain immunity $\omega$ are also unknown.  Therefore we consider model scenarios where these parameters are set at values across their entire range: $0\le\omega\le\sigma\le1$.

Finally, we set the period of time, $u$, over which strain diversity is calculated to be 10 years, to be consistent with the calculation of GAS strain diversity in~\cite{smeesters2009differences}.  

\subsubsection*{Alternative model of the immune response}
In an alternate model of the immune response, we also allow cross-strain immunity to reduce the duration of particular infections by strains that are present in a host when the host clears another strain (so that their new mean duration of infection is $(1-\omega)/\gamma)$.  This effect is only relevant to co-infected hosts.  The parameter exploration described in the main manuscript is repeated with this additional effect of cross immunity, the results of which are provided in electronic supplementary material, Fig.~\ref{fig:figs6}--\ref{fig:figs7}.

\subsection*{Details of model calibration to local epidemiological data}
To calibrate our model to local epidemiological data we used the Python implementation of the Likelihood-Free Inference by Ratio Estimation approach, PYLFIRE~\cite{kokko2019pylfire}, and used a \textsc{Matlab} engine to call our model simulator.  

PYLFIRE is a likelihood-free approach to inferring simulator-based model parameters based on observed data~\cite{kokko2019pylfire}. Similar to other likelihood-free inference approaches, such as Approximate Bayesian Computation (ABC), it involves (1) sampling prior distributions of unknown parameters; (2) generating synthetic data from a simulator-based model using these parameter samples; and (3) calculating the discrepancy between the synthetic and observed data, with the aim of finding regions in the parameter space that result in the smallest discrepancies between the generated and observed data.  

Unlike traditional ABC methods, PYLFIRE uses Bayesian Optimization to efficiently search the parameter space. It infers a regression model that describes the relationship between the discrepancies and the unknown parameters (this relationship is typically modelled as a Gaussian process), which can be used to approximate the likelihood function --  the probability of the observed data given the model parameters~\cite{gutmann2016bayesian}.  Sampling from this estimated likelihood provides an estimate of the normalised posterior distributions of the unknown parameters. Another key difference between ABC and PYLFIRE is that the latter enables the automatic selection of summary statistics that are relevant for inferring model parameters from a set of candidates. It determines the extent to which each summary statistic contributes to the approximation of the likelihood, thus removing the need for this to be user-defined as it is in ABC~\cite{dutta2016likelihood}.   

To run PYLFIRE, we needed to specify the maximum number of requested evidence points $n_e$ where the approximate likelihood function is to be evaluated, the number of initial evidence points $n_0$ that are sampled straight from the prior distributions before starting the Bayesian optimisation process, the number of simulations to run for each evidence point in the parameter space $n_\theta$, and the total number of samples taken from the marginal density distribution $n_m$.  We set ${n_e=300}$, $n_0=50$, ${n_m=320}$ and ${n_\theta=80}$. To estimate the normalized posterior distribution we used a No-U-Turn Markov chain Monte Carlo sampler~\cite{hoffman2014no} to acquire $10^5$ samples from the Gaussian process regression function.

We specified eleven candidate summary statistics to summarise the real and simulated data in a consistent way, including the mean and variance of the monthly point prevalence and point strain diversity, the total number of strains observed over the 23 months, the mean and maximum number of strains observed at a single time point, the maximum number of observations of a single strain at a single time point, the mean and variance of the number of months between observations of particular strains that appeared and reappeared in the data at least once, and the mean normalised point-wise mutual information of the frequency of individual strains at each time point. 

To calculate the summary statistics from the simulated data, we first simulated transmission until the dynamics reached equilibrium (for 30 years) and then continued the simulations for a further 22 months, sampling the data monthly in a manner reflective of the previous study's surveillance protocol, which had significant month-to-month variation in the number of patient consultations~\cite{mcdonald2008dynamic}.  Specifically, 548 people were enrolled in the study in this community but the number of consultations each month ranged from 21 to 211.  Therefore, for each model realisation we assigned 548 hosts uniformly at random from the whole population into the study, and from this pool of hosts, each month we sampled, uniformly at random, the same number of hosts that were seen in the corresponding month of the study.

Our choices of prior distributions for the three unknown parameters $\mathcal{R}_0$, $\sigma$ and $1/\theta$ were informed by the results of the parameter exploration. We assumed the uniform priors ${U}(0,5)$ and ${U}(0.5,1)$  for $\mathcal{R}_0$ and $\sigma$, respectively. For $1/\theta$, we trialled two broad uniform priors: (1) ${U}(5,15)$ years; and (2) ${U}(15,25)$ years. However we did not expect $1/\theta$ to be identifiable in these ranges given the insensitivity of the epidemiological dynamics to large values of $1/\theta$ (see Fig.~2E and SI Appendix, Fig.~\ref{fig:figs4}C).  This was confirmed when we attempted to fit the model to synthetic data generated with $1/\theta=10$ years and with $1/\theta=20$ years (SI Appendix, Fig.~\ref{fig:figs9}--\ref{fig:figs10}).

\subsection*{Parameter values used in figures in the main manuscript}

\subsubsection*{Figure 2} ${1/\gamma=2}$ weeks, ${\beta=0.0301}$, ${n_{\rm max}=42}$, $\alpha=1.71\times10^{-4}$, ${x=10}$. (A--D, F) $c=4.93$; (E) $c=2.39, 3.59,\hdots, 25.05$.

\subsubsection*{Figure 3} ${1/\gamma=2}$ weeks, ${c=4.93}$, ${n_{\rm max}=42}$, $\alpha=1.71\times10^{-4}$, ${\omega=0.1}$, ${N=2500}$, ${x=10}$.

\newpage

\begin{figure}
\centering
\includegraphics[width=\linewidth]{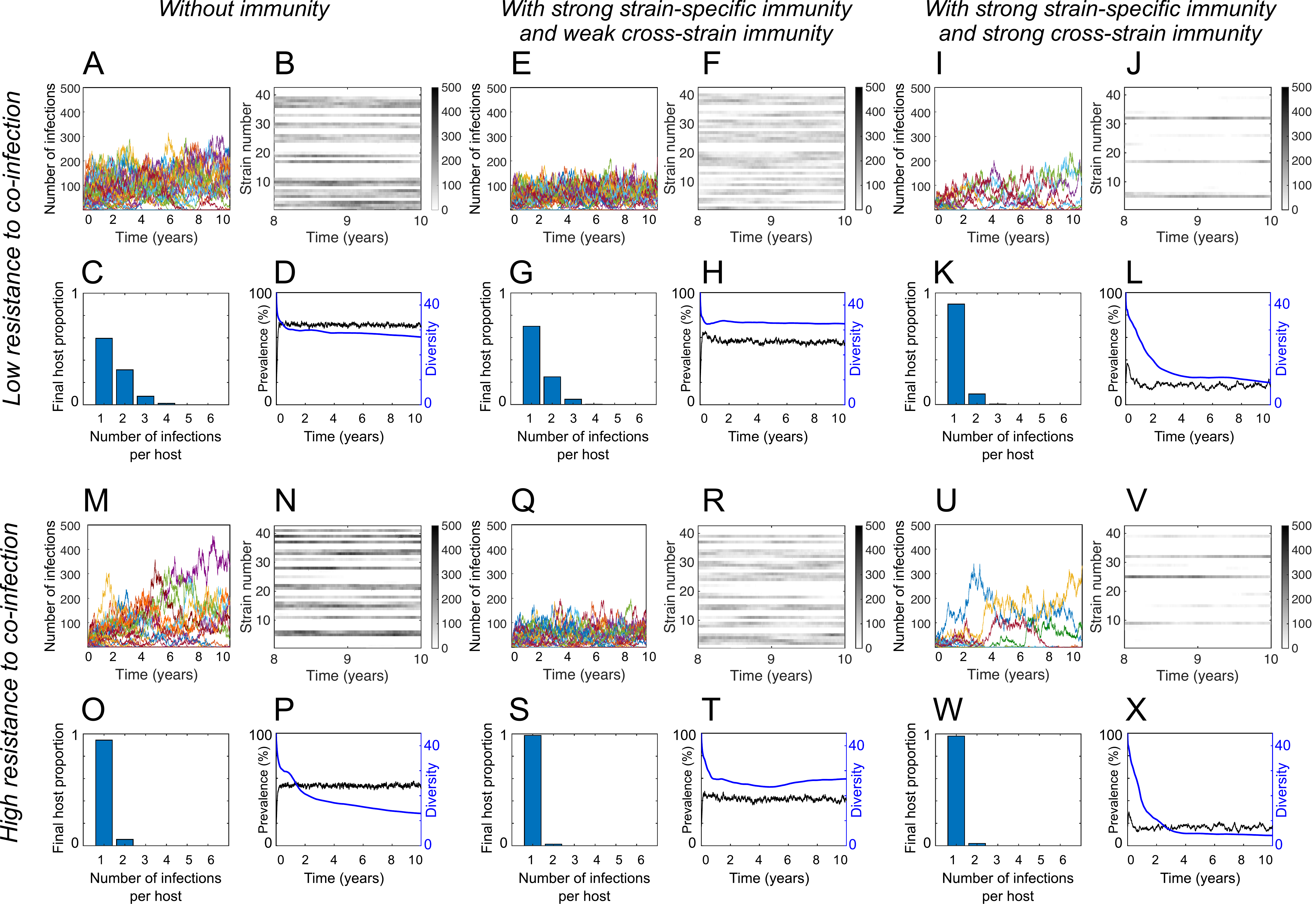}
\caption{Single simulation runs of the model for scenarios where there is (A--L) low; or (M--X) high resistance to co-infection, and (A--D, M--P) no immunity conferred by clearance of infection ($SIS$ model); or temporary immunity is conferred by the clearance of infection ($SIRS$ model) with a high strength of strain-specific immunity $\sigma$ and a (E--H, Q--T) low; or (I--L, U--X) high, strength of cross-strain immunity $\omega$.  (A,E,I,M,Q,R) The number of infections of each strain in the host population (vertical axis) at times $t\in[0,10]$ years (horizontal axis), and the corresponding (B,F,J,N,R,V) number of infections of each strain in the host population in the final two years of each simulation (here, rows correspond to a particular strain, and the shading represents the number of infections), (C,G,K,O,S,W) final distribution of hosts with a given number of co-infections, and (D,H,L,P,T,X) the total prevalence of infected hosts $P(t)$ (left vertical axis and black line) and strain diversity $D(t)$  (right vertical axis and blue line) for time $t\in[0,10]$ years. Here, ${1/\gamma=2}$ weeks, $c=4.93$, ${\mathcal{R}_0=2.07}$, ${n_{\rm max}=42}$, ${N=2500}$, $\alpha=1.71\times10^{-4}$, ${1/\theta=6}$ months, (A--L) ${x=10}$; (M--X) ${x=100}$, and (A--D, M--P) ${\sigma=\omega=0}$; (E--H, Q--T) ${\sigma=0.5}$, ${\omega=0.1}$; and (I--L, U--X)  ${\sigma=0.5}$, ${\omega=0.5}$.  Analogous outputs for the $SIRIR$ model are shown in~\cite{chisholm2019epidemiological}.}
\label{fig:figs1}
\end{figure}

\newpage

\begin{figure}
\centering
\includegraphics[width=0.7\linewidth]{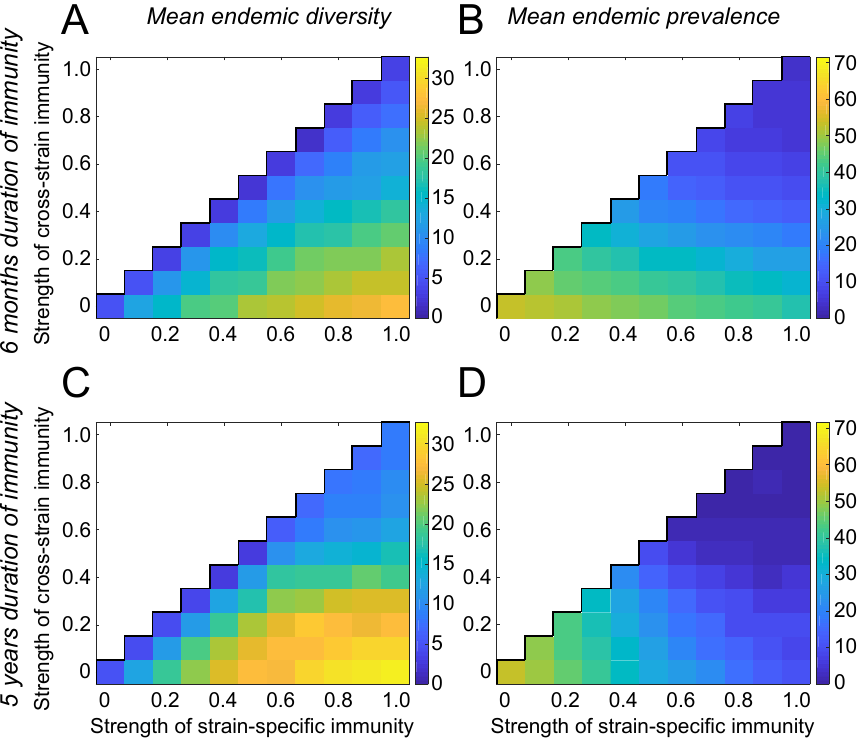}
\caption{The mean endemic (A,C) diversity of strains $D(t)$, and (B,D) prevalence of infections $P(t)$ for different strengths of strain-specific immunity $\sigma$ (horizontal axes) and cross-strain immunity $\omega\le\sigma$ (vertical axes) when there is high resistance to co-infection (${x=100}$). All other parameters match those used in the corresponding scenarios presented in Figure~2A--D.}
\label{fig:figs2}
\end{figure}

\newpage

\begin{figure}
\centering
\includegraphics[width=0.7\linewidth]{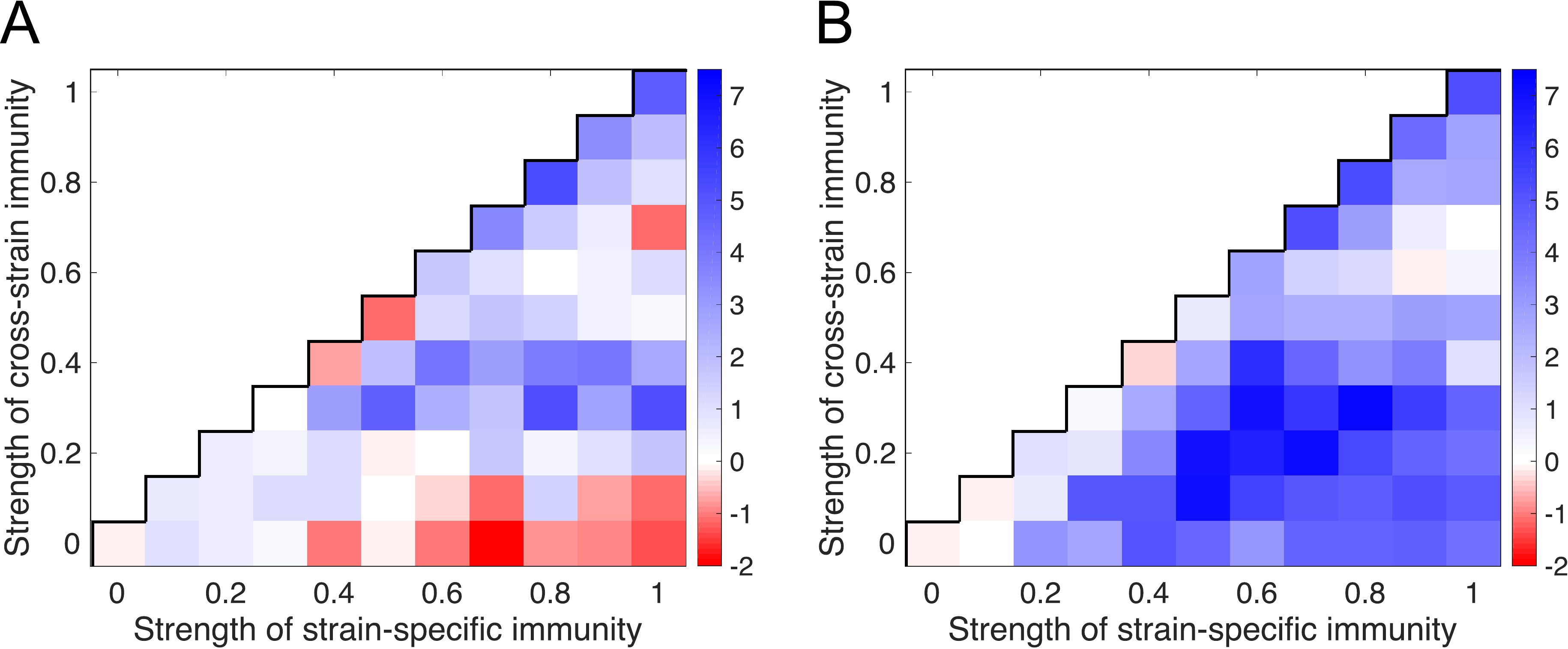}
\caption{The difference between the mean endemic strain diversity when the mean duration of immunity, $1/\theta$, is 5 years compared to when it is 6 months, when there is (A) low resistance to co-infection ($x=10$); and (B) high resistance to co-infection ($x=100$).  All other parameters match those used in Figure~2A--D. Red colours correspond to regions in the parameter space where there is lower endemic strain diversity when the duration of immunity is longer, and blue colours to regions where there is higher endemic strain diversity.  }
\label{fig:figs3}
\end{figure}

\newpage

\begin{figure}
\centering
\includegraphics[width=\linewidth]{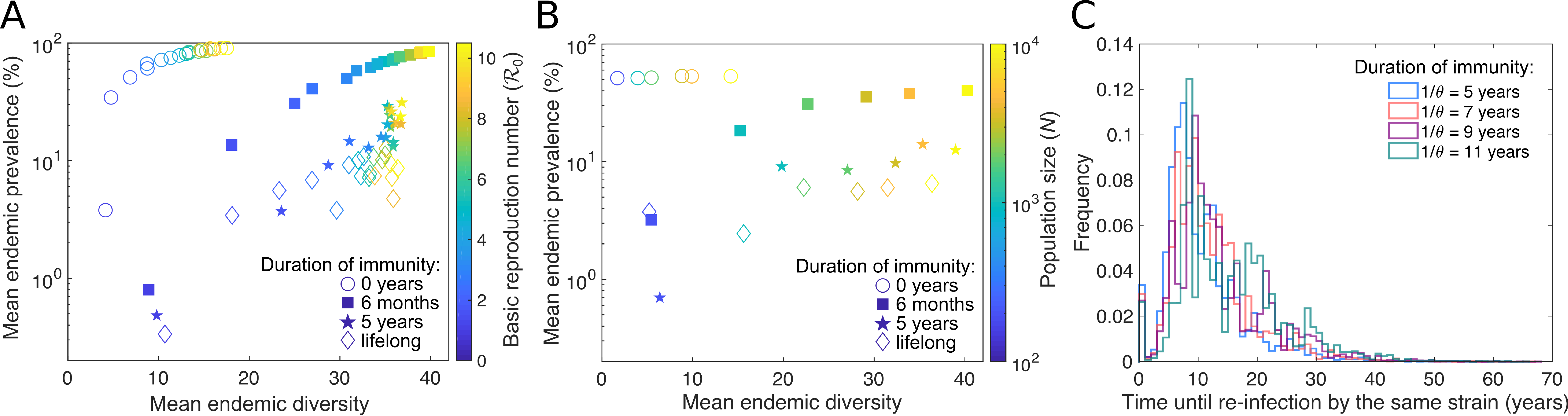}
\caption{Comparing epidemiological dynamics across populations characterised by different values (as indicated by the colour of points) of (A) the basic reproduction number $\mathcal{R}_0$; and (B) the population size $N$, and under different assumptions about $1/\theta$ (indicated by the shape of points) when there is high $\sigma$, low $\omega$ and high resistance to co-infection ($x=100$). The position of points indicates a population's mean endemic $P(t)$ (vertical axes) and associated mean endemic $D(t)$ (horizontal axes) at time ${t=30}$ years (calculated from 10 simulations of the model with a particular parameter combination).  All other parameters match those used in Figure~3E--F. (C) Distributions of the time until re-infection by the same strain for different mean durations of immunity $1/\theta$, calculated from 16 simulations of the model that were run for 100 years for each value of $1/\theta$ considered.  The time between the clearance of a strain and re-infection by the same strain was recorded for each repeat infection, and frequency distributions were generated from the last two years of this data to ensure they represented transmission at dynamic equilibrium.  Here, ${1/\theta=\{5,7,9,11\}}$ years and $\mathcal{R}_0=2$. All other parameters match those used in Figure~2E.}
\label{fig:figs4}
\end{figure}

\newpage

\begin{figure}
\centering
\includegraphics[width=0.7\linewidth]{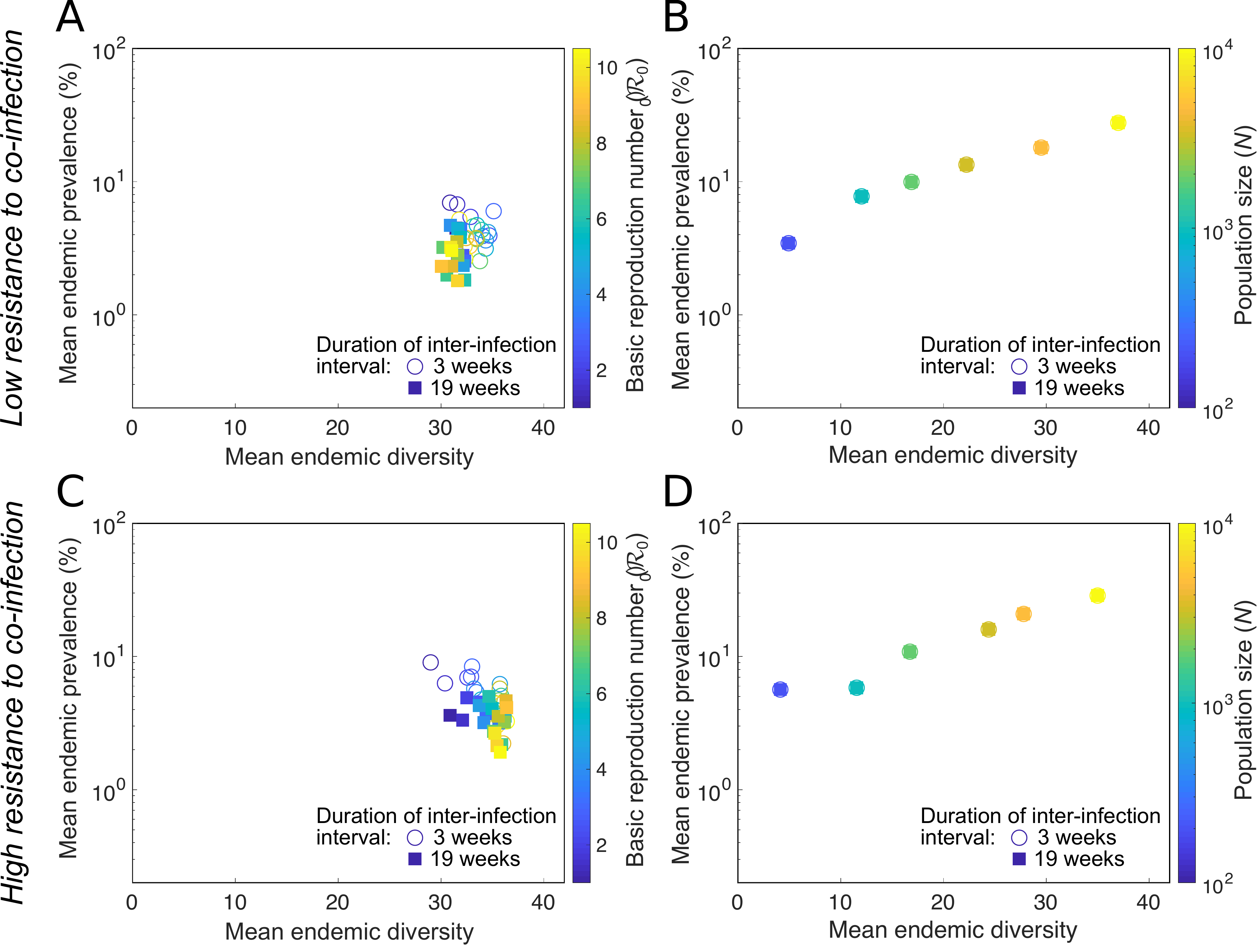}
\caption{Using the $SIRIR$ model to compare epidemiological dynamics in populations characterised by different values (as indicated by the colour of points) of (A,C) the basic reproduction number $\mathcal{R}_0$; and (B,D) the population size $N$, and under different assumptions about the inter-infection interval (indicated by the shape of points) when there is (A,B) low  resistance to co-infection ($x=10$) and (A,B) high  resistance to co-infection ($x=100$). The position of points indicates a population's mean endemic $P(t)$ (vertical axes) and associated mean endemic $D(t)$ (horizontal axes) at time ${t=30}$ years (calculated from 10 simulations of the model with a particular parameter combination).  All other parameters match those used in Figure~2E--F.}
\label{fig:figs5}
\end{figure}

\newpage

\begin{figure}
\centering
\includegraphics[width=\linewidth]{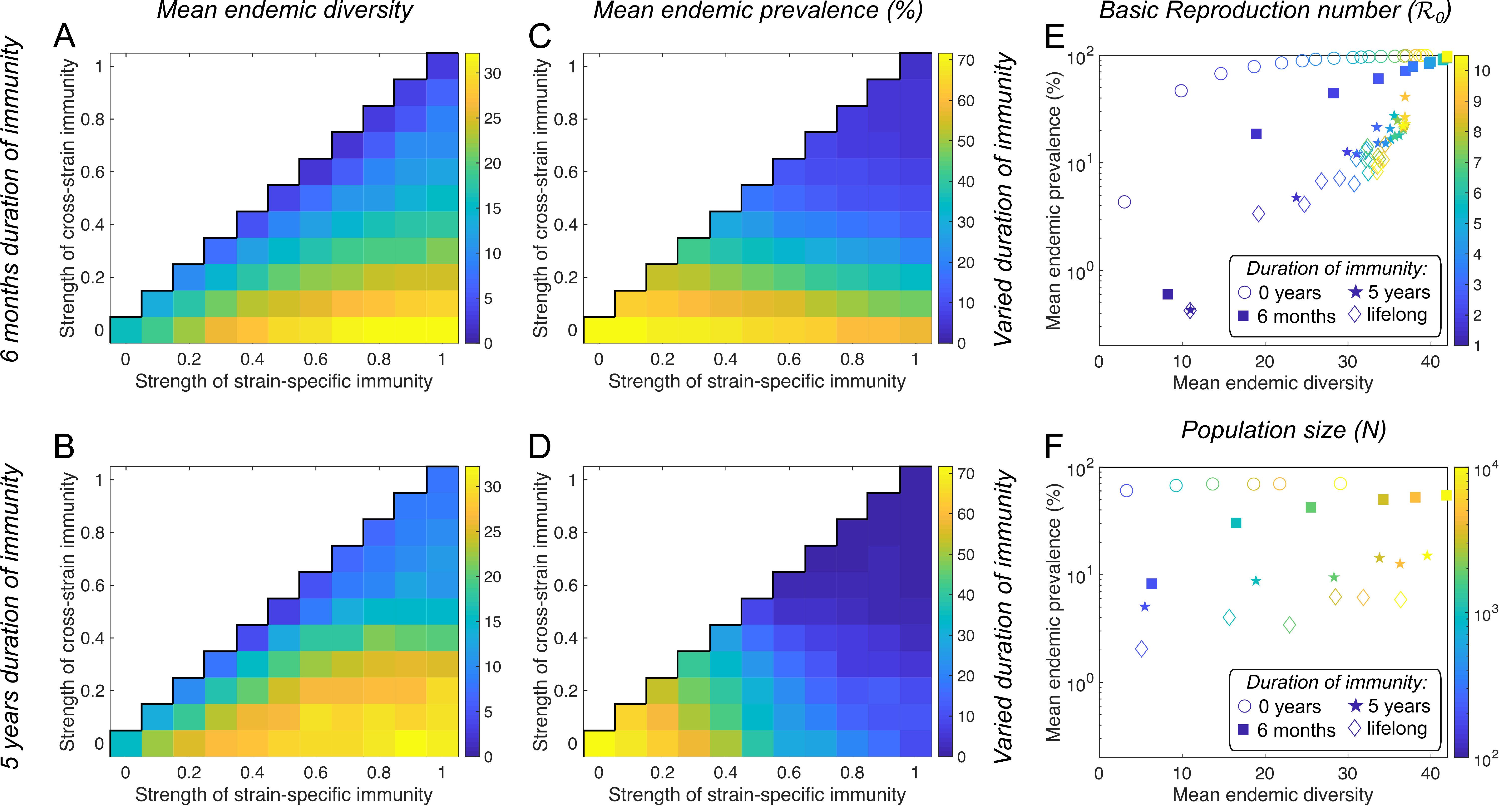}
\caption{Parameter exploration using the alternate model (with extra effect of cross-immunity). The mean endemic (A,B) diversity of strains  $D(t)$, and (C,D) prevalence of infections $P(t)$ for different strengths of strain-specific immunity $\sigma$ (horizontal axes) and cross-strain immunity $\omega\le\sigma$ (vertical axes), when there is low resistance to co-infection, and the mean duration of immunity $1/\theta$ is (A,C) 6 months; and (B,D) 5 years.  Comparing epidemiological dynamics across populations characterised by different values (as indicated by the colour of points) of (E) the basic reproduction number $\mathcal{R}_0$; and (F) the population size $N$, and under different assumptions about $1/\theta$ (indicated by the shape of points) when there is a high $\sigma$, low $\omega$ and low resistance to co-infection. The position of points indicates a population's mean endemic $P(t)$ (vertical axes) and associated mean endemic $D(t)$ (horizontal axes) at time ${t=30}$ years. In all panels, points correspond to the mean of 10 simulations of the model with a particular parameter combination.  Parameters match those used in Figure 2.}
\label{fig:figs6}
\end{figure}

\newpage

\begin{figure}
\centering
\includegraphics[width=\linewidth]{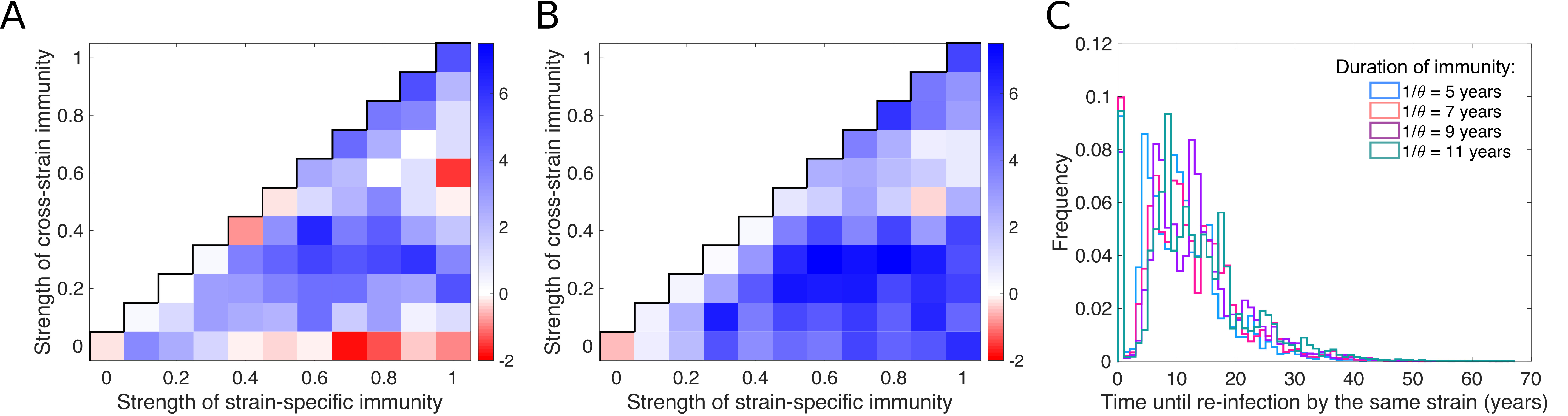}
\caption{Parameter exploration using the alternate model (with extra effect of cross-immunity). The difference between the mean endemic strain diversity when the mean duration of immunity, $1/\theta$, is 5 years compared to when it is 6 months, when there is (A) low resistance to co-infection; and (B) high resistance to co-infection.  Parameters match those used in Figure~\ref{fig:figs3}. (C) Distributions of the time until re-infection by the same strain for different mean durations of immunity $1/\theta$, calculated from 16 simulations of the model that were run for 100 years for each value of $1/\theta$ considered.  Parameters match those used in Figure~\ref{fig:figs4}C.}
\label{fig:figs7}
\end{figure}

\newpage

\begin{figure}
\centering
\includegraphics[width=\linewidth]{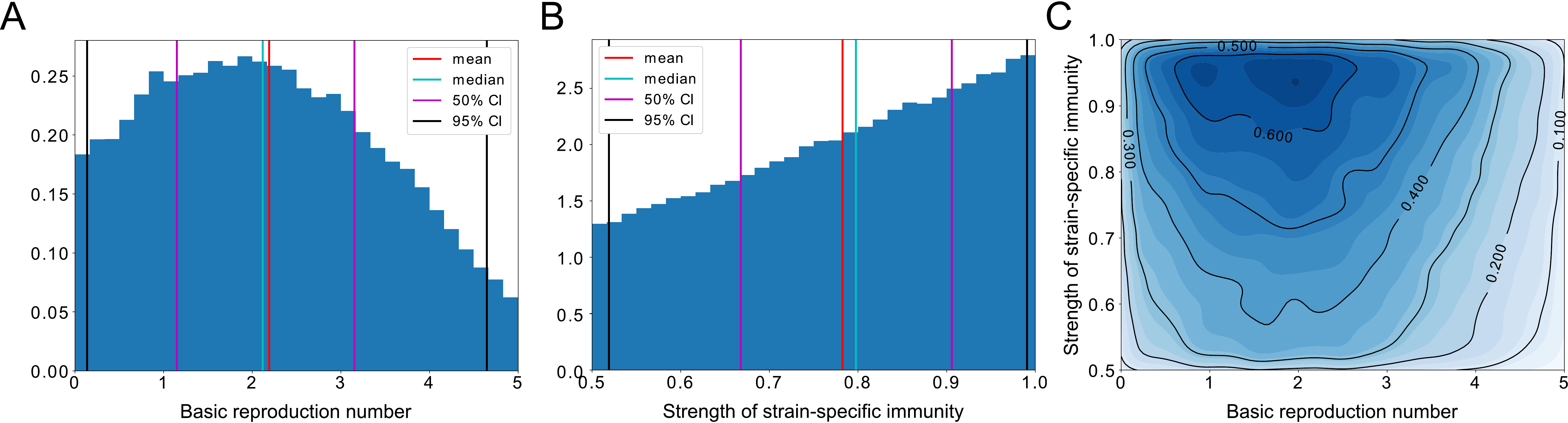}
\caption{Model correlation to Indigenous Australian population assuming mean duration of immunity is between 15--25 years. Estimated (A) marginal distribution for $\mathcal{R}_0$; (B) marginal distribution for $\sigma$; and (C) joint-marginal distribution for $\mathcal{R}_0$ and $\sigma$. All other parameters match those used in Figure 3}
\label{fig:figs8}
\end{figure}

\newpage

\begin{figure}
\centering
\includegraphics[width=\linewidth]{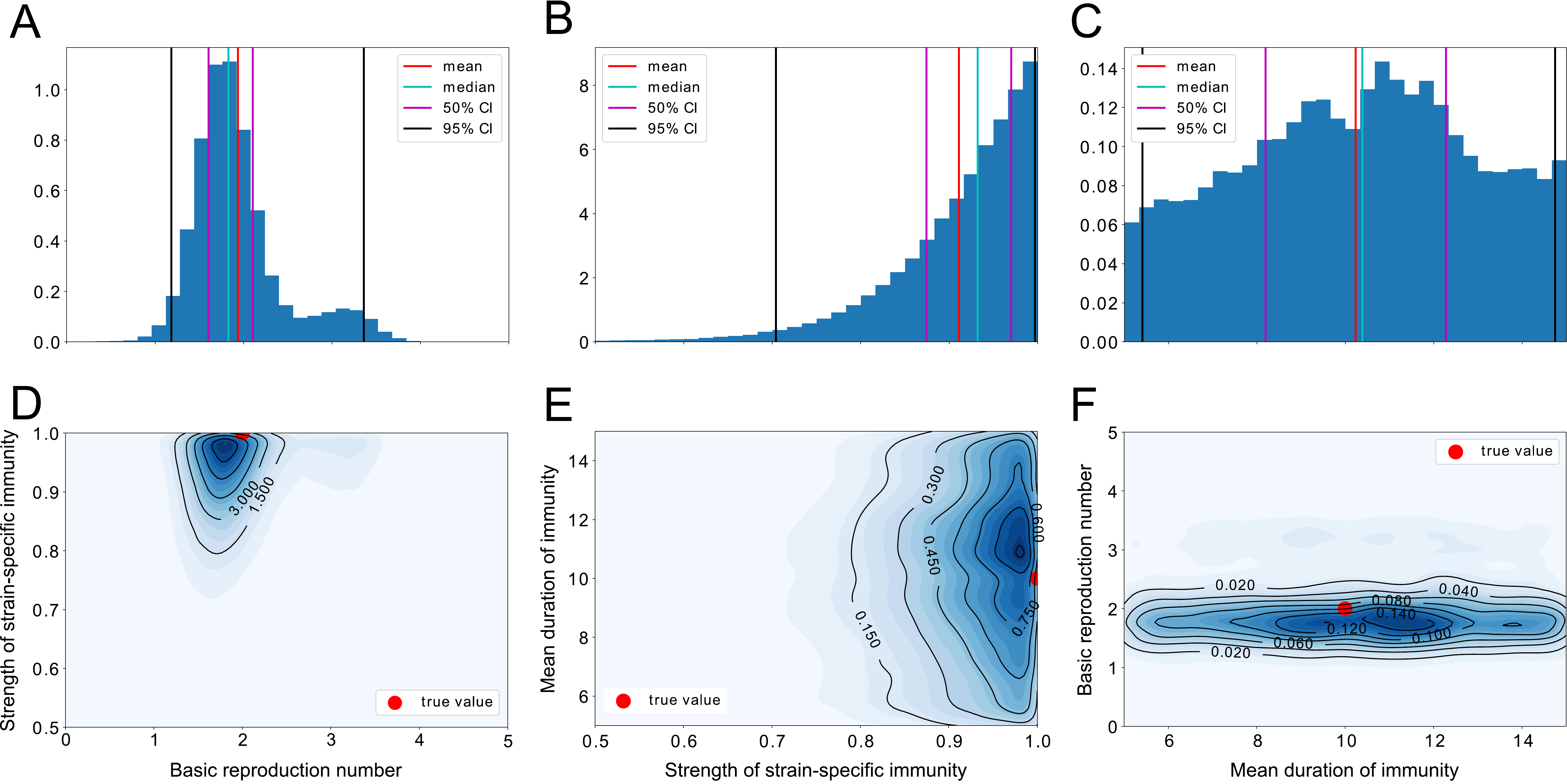}
\caption{Model correlation to synthetic data generated with $\mathcal{R}_0=2$, $\sigma=1$ and $1/\theta=10$ years. Estimated (A) marginal distribution for $\mathcal{R}_0$; (B) marginal distribution for $\sigma$; (C) marginal distribution for $1/\theta$; (D) joint-marginal distribution for $\mathcal{R}_0$ and $\sigma$; (E) joint-marginal distribution for $\sigma$ and $1/\theta$; (F) joint-marginal distribution for $1/\theta$ and $\mathcal{R}_0$.  All other parameters match those used in Figure 3.}
\label{fig:figs9}
\end{figure}

\newpage

\begin{figure}
\centering
\includegraphics[width=\linewidth]{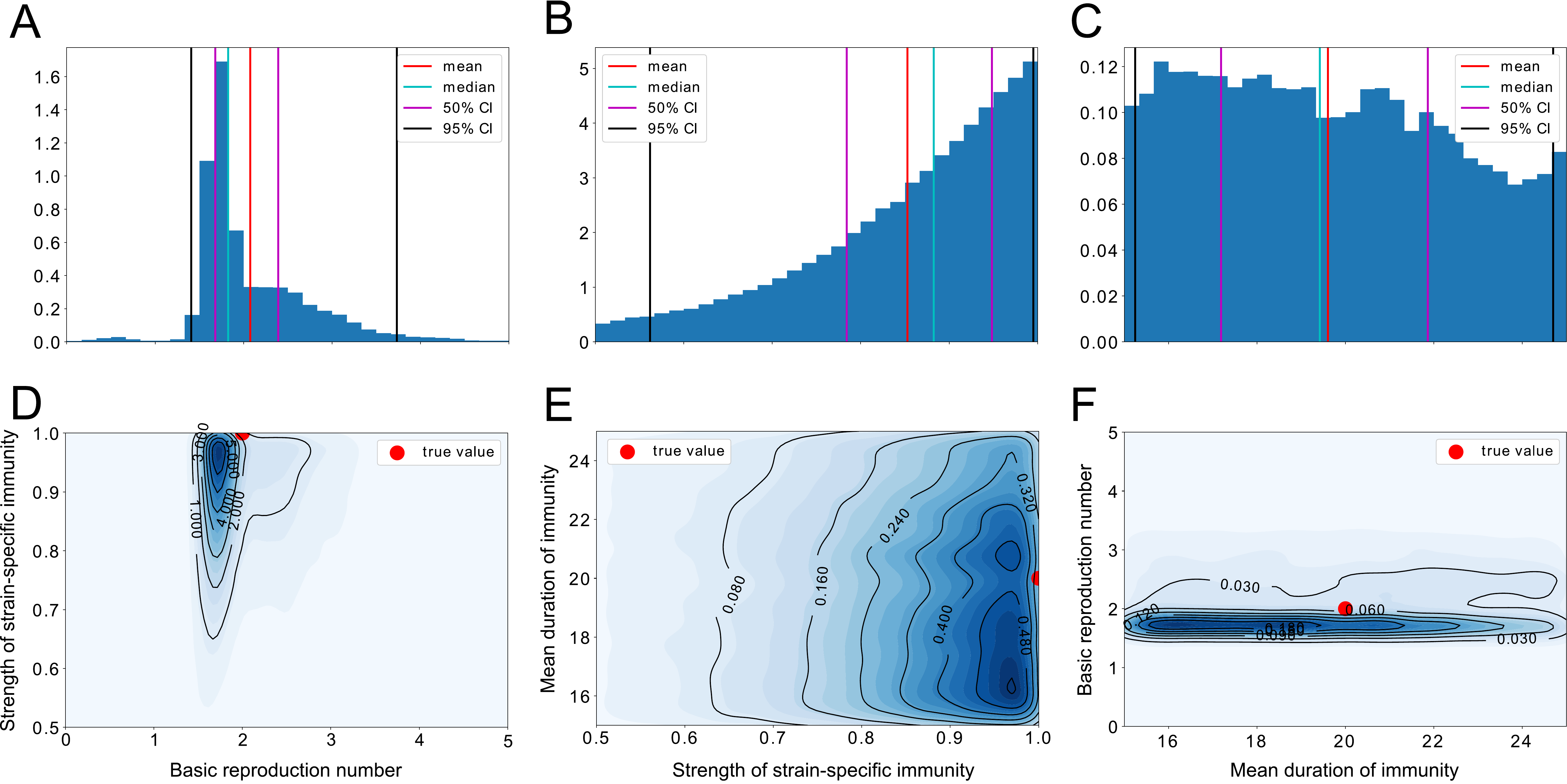}
\caption{Model correlation to synthetic data generated with $\mathcal{R}_0=2$, $\sigma=1$ and $1/\theta=20$ years. Estimated (A) marginal distribution for $\mathcal{R}_0$; (B) marginal distribution for $\sigma$; (C) marginal distribution for $1/\theta$; (D) joint-marginal distribution for $\mathcal{R}_0$ and $\sigma$; (E) joint-marginal distribution for $\sigma$ and $1/\theta$; (F) joint-marginal distribution for $1/\theta$ and $\mathcal{R}_0$.  All other parameters match those used in Figure 3.}
\label{fig:figs10}
\end{figure}

\newpage

\begin{table}[tbhp]

 \caption{Base model parameter values, and alternative formulations}
  \begin{center}
 \label{table:parameters}
\begin{tabular}{lll}
  Parameter  & Base value  & Other values explored \\
\hline
  $N$ & 2500 & 200,1000,2000,3500,5000,10000 \\
  $d$ &  1/71 per year&-\\
  $\alpha$  & 1.71$\times$10$^{-4}$ per day&  -\\
$c$  & 4.93 contacts per day & 2.39,3.59,$\hdots$,25.05 \\
$n_{\rm max}$, $\kappa$  & 42 & - \\
    $\mathcal{R}_0$ & 2.07 & 1,1.5,2,$\hdots$,10.5   \\
    $1/\gamma$ &  2 weeks & - \\
$x$  &-&10, 100\\
$1/\theta$  & - & 0, 0.5, 5, 71 (years) \\
$\sigma$, $\omega$  &-&0,0.1,0.2,$\hdots$,1\\
$w$  & - & 3,19 weeks ($SIRIR$ model)\\
\hline
\end{tabular}
\end{center}
\end{table}

\begin{table}[tbhp]
 \caption{ Bayesian estimates for $\mathcal{R}_0$ and $\sigma$ in an Indigenous Australian population}
 \label{table:fitting}
 \begin{center}
\begin{tabular}{ccccccc}
 Parameter & Prior for $1/\theta$ &  Mean & Median  & Mode & 50\% CI & 95\% CI  \\
\hline
 \centering $\mathcal{R}_0$ & $U[5,15]$ years & 2.27& 2.22&2.17&$(1.27,3.22)$&$(0.18,4.66)$\\
\centering $\mathcal{R}_0$ & $U[15,25]$ years &  2.19&2.12&1.96&$(1.15,3.15)$&$(0.14,4.65)$\\
\centering  $\sigma$ &   $U[5,15]$ years & 0.78&0.79&0.95&$(0.66,0.90)$&$(0.52,0.99)$\\
\centering $\sigma$ & $U[15,25]$ years & 0.78&0.80&0.93&$(0.67,0.91)$&$(0.52,0.99)$\\
\hline
\end{tabular}
\end{center}
\end{table}

\end{document}